\documentclass[10pt]{iopart}
\usepackage{pifont}
\usepackage{graphicx}
\newcommand{\bfsymbol}[1]{\mbox{\boldmath $ #1\!$}}
\begin{document}

\title[Ab initio study of canted magnetism]{Ab initio study of 
canted magnetism of finite atomic chains at surfaces}

\author{B. Lazarovits\dag, B. \'Ujfalussy\ddag, L. Szunyogh\dag \S
\footnote[1]{To whom correspondence should be addressed.
E-mail: szunyogh@heisenberg.phy.bme.hu}, 
G. M. Stocks\ddag \ and P. Weinberger\dag}

\address{\dag\ Center for Computational Materials Science,
Technical University Vienna,
A-1060, Gumpendorferstr. 1.a., Vienna, Austria}

\address{\ddag\
Metals and Ceramics Division, Oak Ridge National Laboratory,
Oak Ridge, Tennessee 37831, USA }

\address{\S\ Department of Theoretical Physics and Center for Applied
Mathematics and Computational Physics,
Budapest University of Technology and Economics,
Budafoki \'ut 8, H-1111, Budapest, Hungary}

\begin{abstract}
By using ab initio methods on different levels we study the magnetic
ground state of (finite) atomic wires deposited
on metallic surfaces. A phenomenological model based
on symmetry arguments
suggests that the magnetization of a ferromagnetic wire is aligned either 
normal to the wire and, generally, tilted with respect to the surface
normal or parallel to the wire. 
From a first principles point of view, this simple model can 
be best related to the so--called magnetic force theorem calculations 
being often used to explore magnetic anisotropy energies 
of bulk and surface systems.
The second theoretical approach we use to search for the canted magnetic 
ground state is first principles adiabatic spin
dynamics extended to the case of fully relativistic 
electron scattering.
First, for the case of two adjacent Fe atoms an a Cu(111) surface
we demonstrate that the reduction of the surface symmetry can indeed
lead to canted magnetism. The anisotropy constants and consequently
the ground state magnetization direction are very sensitive to the
position of the dimer with respect to the surface.
We also performed calculations for a seven--atom Co chain 
placed along a step edge
of a Pt(111) surface. As far as the ground state spin orientation
is concerned we obtain excellent agreement with experiment.
Moreover, the magnetic ground state turns out to be slightly
noncollinear.
\end{abstract}

\pacs{75.10.Lp, 75.30.Gw, 75.75.+a}


\section{Introduction}
\label{sec:intro}

Magnetic devices on the atomic scale became recently a subject of
intensive 
experimental and theoretical research (see, e.g., the ''viewpoint''
drawn by K\"ubler \cite{kubler}). 
Understanding and design of the relevant physical properties --
magnetic moments, magnetic anisotropy energies, thermal stability,
switching -- of atomic scaled magnets demand a detailed knowledge
of their electronic and magnetic structure. For this reason
a considerable amount of theoretical work 
has been published to investigate the -- mostly noncollinear --
magnetic ground state
of free and supported metallic clusters
\cite{ojeda-lopez,oda,ivanov,hobbs,uzdin,fujima,anton}.

Quite recently, Gambardella {\em et al.} reported 
well characterized experiments on linear chains of about 80 Co atoms
located at a step edge
of a Pt(111) surface terrace~\cite{GDM+02}.
At 45 K the formation of
ferromagnetic spin--blocks of about 15 atoms was found with an
easy magnetization axis normal to the chain and pointing along a
direction of 43$^{\rm o}$ towards the step edge.
Stimulated mainly by this experiment, 
in the present work we present a study of the 
magnetic ground state of linear atomic chains deposited on a fcc(111)
host surface. We first focus on the origin of the unusual canted
magnetism. In the case of two adjacent Fe atoms placed into a Cu(111) 
surface we investigate
how the orientation of the magnetization depends on the position 
of the dimer with respect to the surface.
Then we calculate and analyze in some detail 
the ground state spin configuration of a finite 
Co chain at a Pt(111) step edge.

\section{Theoretical methods}
\label{sec:theo}

Let us write the magnetic moment of a ferromagnetic system in terms
of spherical coordinates, $\bfsymbol{M}=M (\sin \theta \cos \phi,
\sin \theta \sin \phi, \cos \theta)$, where $0 \le \theta \le \pi$ and 
$0 \le \phi \le 2\pi$
are the azimuthal and polar angles in a usual rectangular reference of
frame, and assume that the magnitude of the magnetic moment, $M$, is
independent on the the orientation.
In case of a linear chain of atoms deposited 
along the $x$ axis of an fcc(111) lattice (see, e.g.,
figure~\ref{fig:fe2-geometry}) the system has at best
one symmetry operation, namely, a mirror symmetry with respect to the
$(y,z)$ plane, therefore, invariance of the energy 
implies up to second order in the magnetization that
\begin{equation}
\fl \qquad E(\theta,\phi)=E_{0}+K_{2,1}\cos2\theta+ K_{2,2}(1-\cos2\theta)\cos2\phi
+K_{2,3}\sin2\theta\sin\phi \; ,
\label{eq:Ethph}
\end{equation}
\noindent
where $K_{2,i}$ ($i=1,2,3$) are so--called anisotropy constants.
Solving the corresponding Euler--Lagrange equations 
results that, depending on the actual values
of the anisotropy constants, 
the easy magnetization axis corresponds either to $\phi=\pi/2$ and
$\theta \in \{ \theta_0, \theta_0 + \pi/2, \theta_0 + \pi \}$,
where
\begin{equation}
\theta_0 = \frac{1}{2} \arctan \left(\frac{K_{2,3}}{K_{2,1}+K_{2,2}} \right) 
\qquad \left(-\frac{\pi}{4} < \theta_0 < \frac{\pi}{4} \right) \; ,
\label{eq:theta0}
\end{equation}
or to $\theta=\pi/2$ and $\phi=0$.
Clearly, only in the special case of $K_{2,3}=0$ can the ground state 
magnetization point along the $z$ axis (perpendicular to the 
planes).

The so--called magnetic force theorem (MFT) represents a straightforward 
and relatively simple way to calculate anisotropy constants 
as based on the local spin density approximation (LSDA). 
Here, a self-consistent calculation is carried out
for only one selected orientation of the magnetization. Then, by keeping
these potentials and effective fields fixed, the orientation
of the spin-magnetization (in LSDA parallel to the effective field)
is varied, whereby -- neglecting further self--consistency -- 
only the single--site (band) energy is considered,
see reference~\cite{review} for several
applications of this method in ordered and disordered layered systems.
Note that, in principle, also the magnetic dipole--dipole interaction 
energy has to be added to the bandenergy. 
For the case of small nanostructures the estimated magnitude of this energy 
is, however, by at least one order less 
than that of the bandenergy.  

A numerically efficient tool to search for an equilibrium 
spin arrangement is to trace the time evolution
of the spin moments until a stationary state is achieved.
According to the so--called first principles adiabatic spin--dynamics (SD)
founded by Antropov {\em et al.}
\cite{AKH+96}, for a system
with well--defined local (atomic) moments the evolution of the time dependent
orientational configuration, $\{ {\bf e}_i(t) \}$,
is described by a microscopic, quasi-classical equation of motion,
\begin{equation} \label{eq:EOM}
\frac{d {\bf e}_i}{dt} = \gamma \: {\bf e}_i \times {\bf B}_i^{eff} +
\lambda \left[ {\bf e}_i \times ({\bf e}_i \times {\bf B}_i^{eff}) \right]
\; ,
\end{equation}
\noindent
where ${\bf B}_i^{eff}$ is an effective magnetic field averaged
over cell $i$, $\gamma$ is the gyromagnetic ratio and $\lambda$ is
a damping (Gilbert) parameter. Following the arguments of Stocks {\em et al.} 
\cite{SUW+98,UWS+99} in this equation at any moment of
time the orientational state has to be evaluated 
within a constrained density functional theory (DFT). 
Here a local constraining field, ${\bf B}_i^{con}$ 
ensures the stability of a non-equilibrium orientational state. This
implies that the internal effective field that rotates the spins
in the absence of a constraint and, therefore, has to be used in
equation~(\ref{eq:EOM}) is just the opposite of the constraining field
\cite{SUW+98}.
By merging with the locally selfconsistent multiple scattering (LSMS)
method SD has been applied so far to bulk metals and alloys 
\cite{SUW+98,UWS+99,SSS+02} and, very recently, to interfaces
\cite{USS03}.

In order to deal with exchange splitting and relativistic
scattering on equal theoretical footing we combined
the first principles SD scheme based on constrained
DFT by solving the Kohn--Sham--Dirac equation,
\begin{equation} \label{eq:KSD}
\fl \qquad \left[ c \, \mbox{\boldmath $\alpha$} \! \cdot \! {\bf p} +
\beta m c^2 + V({\bf r}) 
+ \mu_B \beta \, \mbox{\boldmath $\sigma$} \! \cdot \!
\left( {\bf B}^{xc}({\bf r}) +
{\bf B}^{con}({\bf r}) \right) - E \right] \psi({\bf r}) = 0
 \; , 
\end{equation}
\noindent
where $\mbox{\boldmath $\alpha$}$ and $\beta$ are the usual Dirac
matrices, $\mbox{\boldmath $\sigma$}$ are the Pauli matrices,
$V({\bf r})$ stands for the Hartree and the exchange--correlation
potential, while within the local spin density approximation
(LSDA) ${\bf B}^{xc}({\bf r})$ is an exchange field interacting
only with the spin of the electron. 
Equations~(\ref{eq:EOM}) and (\ref{eq:KSD}) form the very basis of a
relativistic {\em spin--only} dynamics, inasmuch no attempt is
made to explicitly trace the time evolution of the orbital
moments. 


In conjunction with both the MFT and the SD
we applied the multiple scattering Green's function
embedded cluster method developed by Lazarovits {\em et al.} \cite{LSW02}.
In here, first a self--consistent calculation is carried
out for the surface system in terms of the relativistic Screened
Korringa--Kohn--Rostoker (SKKR) method \cite{SUW95}, and then the nanostructure is
embedded into this host according to the equation
\begin{equation} \label{eq:embed}
\bfsymbol{\tau}^{c}(\epsilon)
=\bfsymbol{\tau}^{r}(\epsilon)
\left[\mathbf{I}-
\left(\mathbf{t}^{r}(\epsilon)^{-1}-
\mathbf{t}^{c}(\epsilon)^{-1}\right)
\bfsymbol{\tau}^{r}(\epsilon) \right]^{-1}
\; ,
\end{equation}
\noindent
where $\bfsymbol{\tau}^r$($\mathbf{t}^r$) and
$\bfsymbol{\tau}^c$($\mathbf{t}^c$) are site--angular momentum
matrices of the scattering path
operators (single--site {\em t} operators) of the host surface
system and the cluster, respectively, and $\epsilon$ is the energy.
By solving also the corresponding Poisson equation with appropriate boundary
conditions a selfconsistent calculation for the selected cluster
can be performed that takes full account of the environment \cite{LSW02}.
It is important to underline that this description does not rely on  periodic 
boundary conditions applied to the (embedded) cluster. 
In all calculations we used the atomic sphere 
approximation (ASA) and an angular momentum expansion up to $\ell_{max}
=2$.

\section{Two Fe impurities at Cu(111) surfaces}
\label{sec:fe2}

\begin{figure}[ht]
\begin{center}
\includegraphics[width=0.4\textwidth,clip]{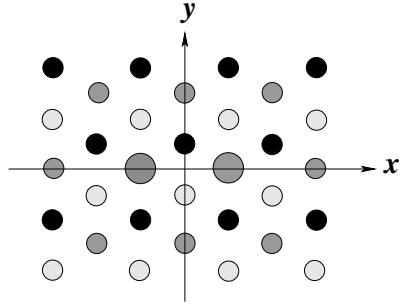} 
\end{center}
\vskip -5 pt
\caption{Schematic top view of two impurities (big shaded circles)
placed into fcc(111) layers as nearest neighbours. The host atoms in the same
layer are displayed by small shaded circles, those in the layer above by
full circles, while those in the layer below by empty circles.}
\label{fig:fe2-geometry}
\end{figure}

The simplest system that reduces the symmetry of the fcc(111) layers to only
one mirror plane is a pair of nearest neighbour defects 
as illustrated in figure \ref{fig:fe2-geometry}.
We embedded two such Fe impurities on top, i.e., into the
first vacuum layer (labelled by $S+1$), into the surface ($S$) layer
and into the subsurface ($S-1$) layer of a Cu(111) surface. 
First a self--consistent SKKR calculation 
has been performed by relaxing the potentials of 6 Cu layers and 3 empty
sphere layers in order to describe the surface region from the bulk to the
vacuum. Then, by employing equation (\ref{eq:embed}), 
for the three above cases
the Fe dimers were calculated self--consistently.
Since these calculations are intended provide only with qualitative
predictions, for simplicity, we neglected relaxations of the potentials 
of the host atoms,

\begin{figure}[hb]
\begin{center}
\includegraphics[width=1.00\textwidth,clip]{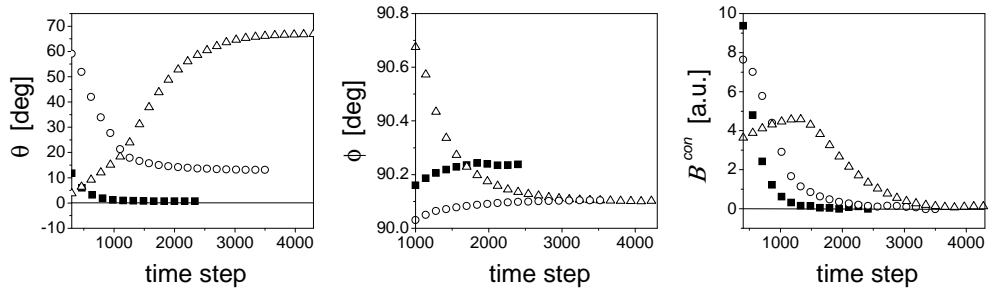}  
\end{center}
\vskip -5 pt
\caption{Evolution of the direction of magnetization and the magnitude
of the constraining
field according to a spin--dynamics calculation for two Fe impurities
placed into different layers of a Cu(111) surface: 
\ding{110} \; $S+1$, $\triangle$ \; $S$, \ding{109} \; $S-1$.
For better visibility, the corresponding results for
the first 300--1000 iterations are not shown. Displayed are only the
data for one of the Fe atoms (see text).
}
\label{fig:fe2-sd}
\end{figure}

As mentioned before, the present implementation of our SD scheme
serves (only) for finding the magnetic ground state of the system,
therefore, it is sufficient to consider only the second
(damping) term on the right hand side 
of equation~(\ref{eq:EOM}). The evolution of the spin orientation is
then measured on a time scale with a unit (time step) of
$1/\lambda$. In figure~\ref{fig:fe2-sd} the evolution of the $\theta$
and $\phi$ angles and the magnitude of the constraining field are plotted 
in this artificial time scale for one of the Fe atoms. 
Note that during the SD procedure, the magnetic configuration 
of the two Fe atoms was confined to be symmetric
with respect to the $(y,z)$ plane. (We checked, however, that the final
result is independent on the starting configuration.)

Actually in a few time steps, for all the three layer positions the
magnetic state of the two Fe atoms became nearly ferromagnetic 
and perpendicular to the line connecting the
two impurities ($\phi \simeq 90^\circ$). A satisfactory convergence was,
however, achieved only after thousands of time steps later, when -- 
as can be seen from figure~\ref{fig:fe2-sd} -- 
the constraining fields converged 
to zero. The final magnetic states can be summarized as follows:
$\theta= 0.73^\circ, 66.8^\circ$ and $13.1^\circ$, as well as
$\phi=89.8^\circ, 89.9^\circ$ and $89.9^\circ$ for one of the Fe atoms
in layers $S+1$, $S$ and $S-1$, respectively, and a symmetric
orientation for the other Fe atom. Clearly, these magnetic ground states
are in qualitative agreement with the predictions of the simple
phenomenological theory, see equations (\ref{eq:Ethph}) and
(\ref{eq:theta0}).

\begin{figure}[htbp]
\begin{center}
\includegraphics[width=0.80\textwidth,clip]{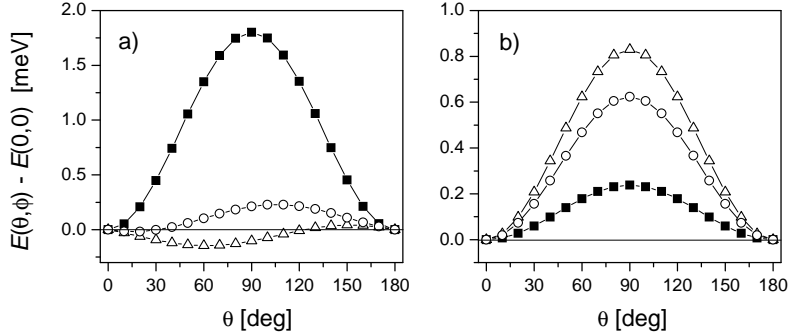}  
\end{center}
\vskip -10pt
\caption{Bandenergy differences from MFT calculations for two Fe
impurities at a Cu(111) surface: \ding{110} \; $S+1$, $\triangle$ \;
$S$, \ding{109} \; $S-1$; a) $\phi=90^\circ$, b) 
$\phi=0$. Solid lines show curves fitted to equation \ref{eq:Ethph}
with the parameters contained by table \ref{tab:fe2-parameters}.}
\label{fig:fe2-fp}
\end{figure}
\vskip -0.3cm
\begin{table}[htbp]
\caption{Anisotropy parameters (in units of meV), see equation
(\ref{eq:Ethph}),
 fitted to energies calculated within the MFT 
for two Fe impurities placed at different layers of a Cu(111) surface.
The azimuthal angle corresponding to the global minimum of the energy
is also displayed.}
\lineup
\begin{indented}
\item[]
\begin{tabular}{cllll} 
\br
layer & $K_{2,1}$ & $K_{2,2}$ & $K_{2,3}$ & $\theta_0$ \\
positions &&&& \\[4pt] \hline \\[-8pt]
$S$+1 &  --0.51 &  --0.39 & --0.0026 & \00.083$^\circ$ \\
$S$   &  --0.18 & \00.23 & --0.079  & 62.0$^\circ$  \\
$S$-1 &  --0.21 & \00.10 & --0.065  & 15.7$^\circ$ 
\\ \br
\end{tabular}
\end{indented}
\label{tab:fe2-parameters}
\end{table}

We carried out MFT calculations by using the output of the SD
calculations.
Figure~\ref{fig:fe2-fp} shows (band)energy curves
when scanning the (uniform) direction of the magnetization through the paths
$0 < \theta < 180^\circ$ for $\phi=90^\circ$, i.e., 
in the $(y,z)$ plane and for $\phi=0^\circ$, i.e., in the $(x,z)$ plane.
All the calculated data fit almost precisely the function in equation
(\ref{eq:Ethph}) with the parameters listed in table
\ref{tab:fe2-parameters}. 
For all the three cases 
the minimum of the energy is found in the $(y,z)$ (symmetry) plane 
at an azimuthal angle, $\theta$, also shown in table \ref{tab:fe2-parameters}. 
These angles coincide remarkably well with those obtained from the SD
calculations.  It should be noticed, however, that by using the output
of self--consistent calculations with a magnetization fixed along 
the $z$ axis, in particular, for an Fe dimer placed into layer $S$
we obtained an apparently different ground state orientation
($\theta=46.5^\circ$). This clearly demonstrates that the applicability
of MFT is quite limited. 

It is obvious that a pair of Fe atoms in layer $S+1$ represents the case
of strong perpendicular anisotropy with a corresponding anisotropy
energy of about 0.9~meV/Fe atom. This 
clearly explains the relatively fast convergence of the SD scheme,
while in the other two cases due to the smaller spin--orbit coupling the
convergence was much slower (see figure \ref{fig:fe2-sd}).
For Fe dimers immersed into
layers $S$ and $S-1$ a change of the orientation from the
$z$ to the $x$ direction costs much larger energy 
than a corresponding variation within the symmetry plane, because
the otherwise remarkably reduced anisotropy constants, $K_{2,1}$ and
$K_{2,2}$, differ in sign. Concomitantly, the relative increase of 
the magnitude of $K_{2,3}$ gives rise to a canted ground state, see
equation~(\ref{eq:theta0}). 
It should be noted that, similar to as discussed in Ref.~\cite{LSWU03},
the change of the anisotropy
constants, $K_{2,i}$ with respect to the position of the Fe dimer can
be related to the different hybridization between the electronic
states of the Fe and the Cu atoms. 

\section{Finite Co wire at the step edge of a Pt(111) surface}
\label{sec:co7}

\begin{figure}[htbp]
\begin{center}
\includegraphics[width=0.90\textwidth,clip]{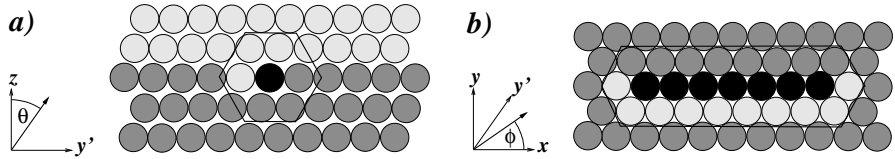} 
\end{center}
\caption{Schematic view of the geometry of a seven--atom Co chain
along a Pt(111) step edge. Full circles: Co atoms,  shaded
circles: Pt atoms, open circles: empty spheres.  $\mbox{\boldmath
$a)$}$: side view, $\mbox{\boldmath $b)$}$:
top view of the surface Pt layer with the Co chain. The
cluster embedded is indicated by solid lines. The coordinate
system gives reference to the azimuthal and
polar angles, $\theta$ and $\phi$, that characterize the
orientation of the magnetization. (Note that in
Ref.~\cite{GDM+02} a different coordinate system and
the opposite notation for the angles is used.)}
\label{fig:geometry}
\end{figure}

We first
performed a calculation treating 8 layers of Pt selfconsistently
together with  4 layers  of vacuum. Then, a seven--atom chain of
Co together with 10 empty (vacuum) spheres were embedded into the
topmost Pt layer as schematically indicated in figure
\ref{fig:geometry}. Simultaneously all the nearest neighbours of
the Co atoms were re--embedded into the respective Pt or vacuum
layers to allow for the relaxation of potentials around the Co chain.
Therefore, an embedded cluster of a total of 55 atoms was treated
selfconsistently. 

\begin{figure}[htbp]
\begin{center}
\begin{tabular}{cc}
\includegraphics[width=0.48\textwidth]{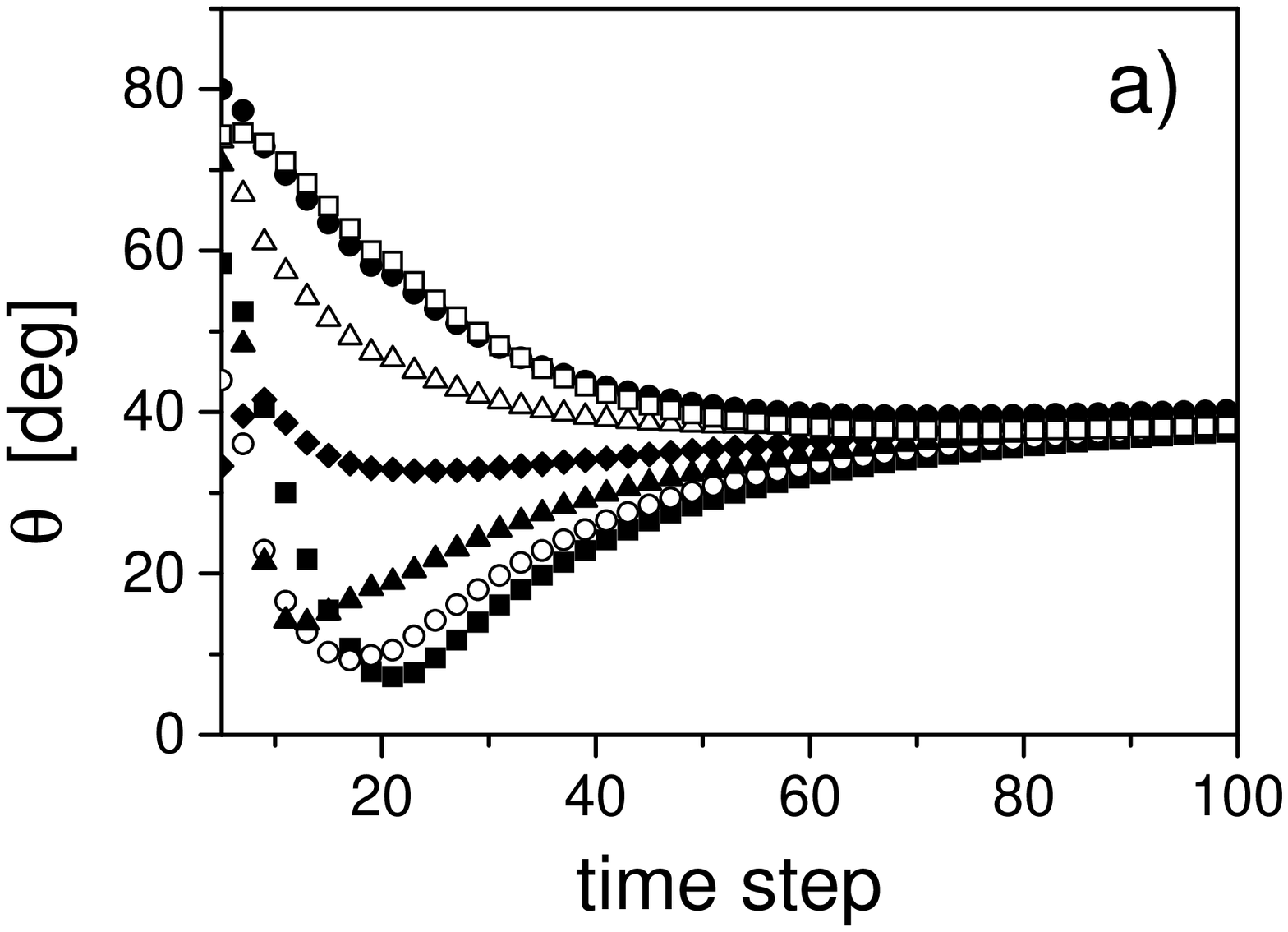}  &
\includegraphics[width=0.48\textwidth]{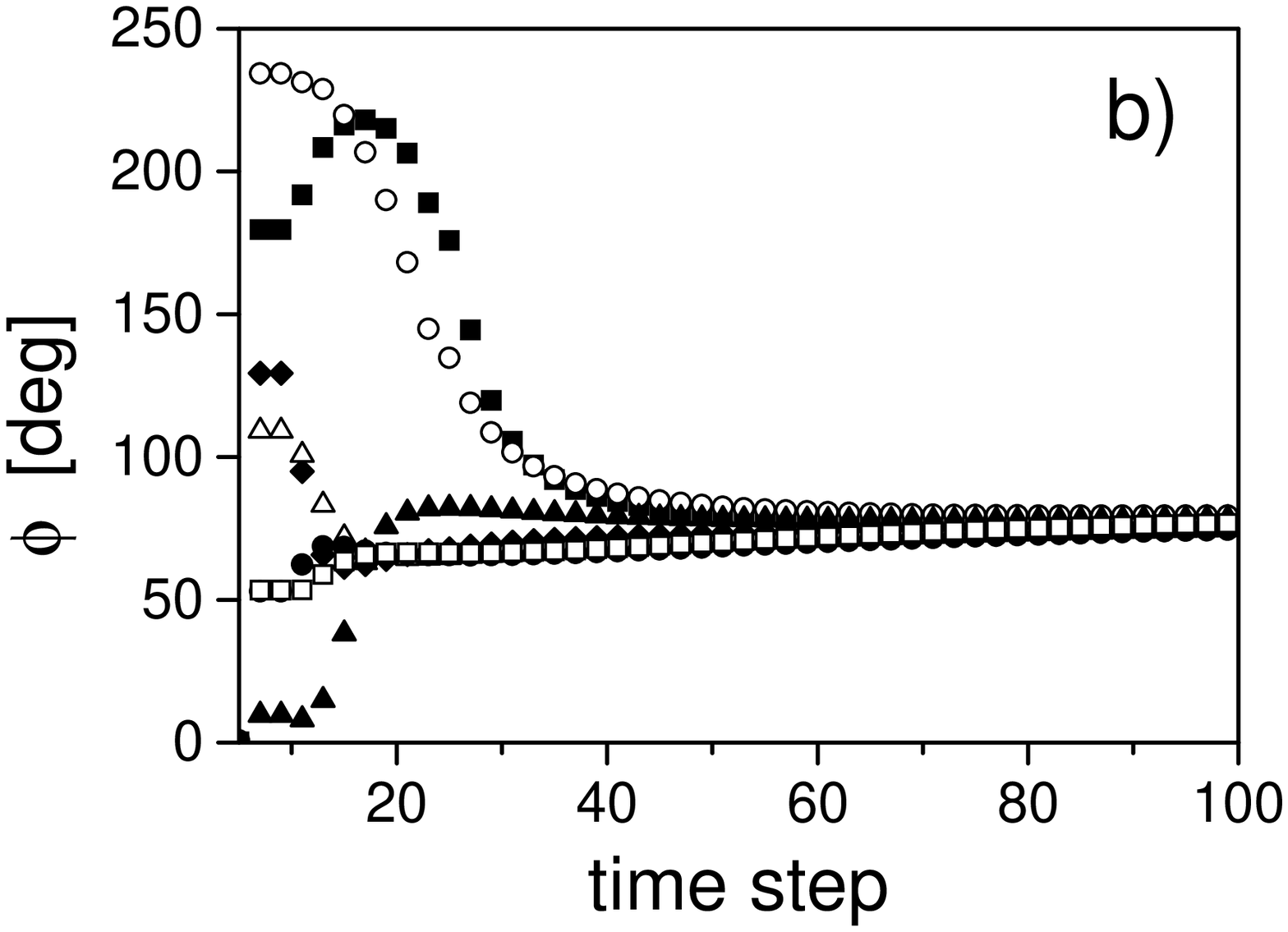} 
\end{tabular}
\end{center}
\vskip -10pt
\caption{Evolution of the angles $\theta$, part a), and $\phi$,
part b), defining the orientation of the spin moments for the
seven Co atoms in the finite chain depicted in
figure~\ref{fig:geometry}. The symbols refer to the following Co
atoms numbered from the left to the right in part b) of
figure~\ref{fig:geometry}: \ding{110}  \ 1, \ding{109} \ 2,
\ding{115} \ 3, \ding{117} \ 4, $\triangle$ \ 5, \ding{108} \ 6,
\ding{112}  \ 7. 
Shown are only the first 100 time steps. 
}
\label{fig:co7-sd}
\end{figure}

For each Co atom in the chain, in figure~\ref{fig:co7-sd} 
the evolution of the $\theta$
and $\phi$ angles is plotted for the first 100 time steps in the
artificial time scale mentioned in \sref{sec:fe2}. Initially the
directions of the atomic magnetic moments were set by a random
number generator. It can be seen that after some oscillations both
the $\theta$ and the $\phi$ angles for all the Co atoms quickly
approach to nearly the same value for all the Co atoms, i.e.,
as in the previous case of Fe dimers at a Cu(111) surface,
to a nearly ferromagnetic configuration.
The initial
rapid oscillations that can be observed in figure~\ref{fig:co7-sd} are
the consequence of the relatively large constraining fields caused
by large exchange energies whenever the moments point into
very different directions. 
In about 1000 time steps the $\phi$ angles, converged to
90$^\circ$, with a spread of less then 1$^\circ$, i.e., normal to
the chain and the $\theta$ angles converged to nearly 42$^\circ$.
All these results are in excellent agreement with
experiment~\cite{GDM+02}. 

We also performed MFT calculations 
with the selfconsistent potentials
and fields obtained from the SD procedure.
The fitted anisotropy parameters $K_{2,1}=-0.16$ meV,
$K_{2,2}=-1.06$ meV, and $K_{2,3}=-4.81$ meV 
not only fairly well reproduce the easy axis, $\theta=38^\circ$ and
$\phi=0^\circ$, but also result in
a value of the anisotropy energy (defined as the energy 
difference between the hard and the easy axes)
of 1.42 meV/Co atom that satisfactorily compares 
with that derived from experiment (2.0 meV) \cite{GDM+02}.

\begin{table}[ht]
\caption{Calculated magnitudes and orientations of the spin and
orbital moments in a seven--atom Co chain along a Pt(111) step edge.}
\begin{indented}
\item[]
\begin{tabular}{ccccc}
\br   
atom & \multicolumn{2}{c}{Spin moment} & \multicolumn{2}{c}{Orbital moment}
\\ 
  & moment($\mu_B$) & $\Theta$($\deg$) & moment($\mu_B$) & $\Theta$($\deg$) 
  \\[3pt] \hline \\[-7pt]
1 & 2.23 & 41.1 & 0.25 & 39.1 \\
2 & 2.18 & 42.5 & 0.20 & 41.5 \\
3 & 2.18 & 42.3 & 0.19 & 40.1 \\
4 & 2.18 & 42.4 & 0.20 & 41.3 \\
5 & 2.18 & 42.3 & 0.19 & 40.2 \\
6 & 2.18 & 42.5 & 0.20 & 41.5 \\
7 & 2.23 & 41.1 & 0.25 & 39.1 \\ \br   
\end{tabular}
\end{indented}
\label{tab:co7}
\end{table}

Extracted from the final (equilibrium) state, the size and the
azimuthal angle $\theta$ of the spin and orbital moments 
for each Co atom are shown
in table~\ref{tab:co7}. While the
calculated spin moments for the inner Co atoms (2.18 $\mu_B$) are
in good agreement with the value deduced from experiment (2.12
$\mu_B$)~\cite{GDM+02} and also with other theoretical studies on
infinite wires \cite{KED+02,SMO03}, 
the edge of the wire is characterized
by larger  spin (and orbital) moments~\cite{LSW03}. 
Although our calculated
orbital moments for the inner atoms (0.19--0.20 $\mu_B$) are
larger than the corresponding values from other LSDA calculations
(0.16 $\mu_B$ \cite{KED+02} and 0.15 $\mu_B$ \cite{SMO03}), they
are still much too small when compared to the experimental value
(0.68 $\mu_B$) \cite{GDM+02}. 
Note that including orbital polarization scheme or using the LDA+U
method a value of 0.92 $\mu_B$ \cite{KED+02} and
0.45 $\mu_B$ \cite{SMO03} can be obtained.

As can be inferred from table~\ref{tab:co7} the spin moments
of the inner atoms are fairly parallel, those at the end of the chain
however, are off by more than 1$^\circ$. This can be associated with the
anisotropy energy contributions being larger at end of the chains
than inside as found for finite Co wires deposited on a Pt(111)
surface \cite{LSW03}. It can also be seen in table~\ref{tab:co7}
that the orbital moments oscillate stronger in magnitude and orientation
than the spin moments.
As pointed out by Jansen \cite{Jan99}, within the DFT the spin
and orbital moments are required to align only when the ground state refers to
a high--symmetry direction. This is, however, not the case for the
Co wire since the ground state orientation is not directly induced by
symmetry.

\section{Summary}
In this work we presented calculations of the magnetic ground
state of linear atomic chains placed onto 
surfaces in terms of the magnetic force theorem and an ab-inito
spin dynamics scheme.
We found that due to the variation of the anisotropy constants
the canted magnetic state obtained for a Fe dimer at
a Cu(111) sensitively depends on its position. 
In excellent quantitative agreement with experiment,
we obtained a canted ground state for a finite
Co wire along a Pt(111) surface step edge.
We also found that this
magnetic state is noncollinear: a feature that is expected to play a key role
in nanostructures of more complex geometry. 

\ack
Financial support was provided by the Center for Computational
Materials Science (Contract No. GZ 45.531), the Austrian Science
Foundation (Contract No. W004), the Research and Technological
Cooperation Project between Austria and Hungary
(Contract No. A-3/03) and the Hungarian National
Scientific Research Foundation (OTKA T046267 and OTKA T037856).
The work of BU and GMS was supported by DOE-OS, BES-DMSE under
contract number DE-AC05-00OR22725 with UT-Battelle LLC.
Calculations were performed at ORNL-CCS (supported by OASCR-MICS)
and NERSC (supported by BES-DMSE).

\end{document}